\documentclass[final,leqno]{siamltex}
\usepackage{graphicx}
\usepackage{amssymb}
\usepackage{amsmath}
\usepackage{float}
\usepackage{graphics,color}
\usepackage{xcolor}

\newcommand{\pd}[2]{\frac{\partial #1}{\partial #2}}

\title{Heat equation and Brownian motion of an overdamped rotating sphere}
\author{H.B. van Lengerich\thanks{Work performed while at the Department of Aerospace Engineering and Mechanics, University of Minnesota, 107 Akerman Hall, 110 Union St. SE, Minneapolis, MN 55455-0153  currently at 3M, Maplewood, MN (\tt henrikvl@gmail.com})}

\begin{document}
\maketitle

\begin{keywords} 
overdamped, rotation, stochastic, Ito, Stratonovich, Brownian
\end{keywords}

\begin{AMS}
58J65, 60J60, 82B31, 82C31
\end{AMS}

\pagestyle{myheadings}
\thispagestyle{plain}
\markboth{H.B. VAN LENGERICH}{HEAT EQUATION AND BROWNIAN MOTION OF ROTATION SPHERE}

\begin{abstract}
The governing equations of Brownian rigid bodies that both translate and rotate are of interest in fields such as self-assembly of proteins, anisotropic colloids, dielectric theory, and liquid crystals. In this paper, the partial differential equation that describes the evolution of concentration is derived from the stochastic differential equation of a sphere experiencing Brownian motion in a viscous medium where a potential field may be present. The potential field may be either interactions between particles or applied externally. The derivation is performed once for particles whose orientation can be specified by a vector ($S^2$), and again for particles which require a rotation matrix ($SO(3)$). The derivation shows the important difference between probability density and concentration, the Ito and Stratonovich calculus, and a Piola-type identity is obtained to complete the derivation.
\end{abstract}

\section{Introduction}
Rotational Brownian motion is important in fields such as self-assembly of viruses \cite{hagan06}, Janus particles \cite{walther13}, dielectric theory \cite{McConnell80} and liquid crystals \cite{freed77}. For example, consider the self-assembly of a virus capsid. Many virus proteins are translating and rotating in a Brownian motion due to collisions with the solvent molecules. The solvent also causes a drag which resists translation and rotation. The drag is very large compared to the change in momentum of the protein, hence the motion of the protein is overdamped. The proteins have bonding sites at various locations, therefore the orientation of the protein is important in modeling the assembly process. For simplicity, spherical proteins will be assumed - the validity of this assumption depends on the particular virus. Given this physical description, we infer that the self-assembly of a virus can be described by the overdamped Brownian motion of spheres that interact in potential field that depends on orientation of the protein.

The properties of Brownian systems can be obtained by numerically simulating trajectories of the overdamped Langevin equation \cite{coffey04, hagan06, frenkel01}. The Langevin equation is a stochastic differential equation (SDE) which represents a force and torque balance that includes random torques and forces due to the collisions with the solvent molecules. The overdamped Langevin equation describing rotational motion is sensitive to the type of random noise - whether it be Ito or Stratonovich. One of the goals of this paper is to describe why the Stratonovich interpretation correctly captures the physical situation. When performing numerical simulations, the Langevin equation can be converted to an equation that uses the Ito noise so that an Euler integration can be used \cite{evans01, higham01}. The necessity for this mathematical care is unusual - the overdamped Langevin equation in Cartesian coordinates and the Langevin equation without the overdamped approximation in any coordinate system do not depend of the type of noise that is used.

It is often desirable to find the ensamble averaged distribution of positions and rotations of particles, these can be expressed as either concentration or probability. For example, if points on a globe are picked randomly such that all area elements on the surface are equally likely to be picked, we say the concentration of points will be uniform. If points on a globe are picked such that they are equally likely to lie between any latitudinal and longitudinal coordinates, we say the probability is constant. In the case where the probability is constant, it is equally likely for points to be between 0 and 10 degrees latitude and 80 and 90 degrees latitude. Because the poles have a smaller area, the concentration of particles will be higher at the poles when the probability is uniform.

It has been known by Perrin \cite{perrin34} that the evolution of the concentration of a rotating overdamped Brownian particle is the heat equation. The main goal of this paper is to derive the evolution equation for the concentration in an arbitrary rotational coordinate system for the Brownian motion of a vector ($S^2$) and for a three dimensional particle ($SO(3)$) starting from the overdamped Langevin equation. 

The equivalence of the two methods is typically assumed; however, for rotating particles it involves three subtle points. First, the overdamped Langevin equation is a Stratonovich stochastic differential equation, and not an Ito SDE. Second, the probability density is different than the concentration. Typically, the Fokker-Planck equation gives the evolution of the probability density and the heat equation gives the evolution of concentration. Finally, an additional identity, similar to Piola's identity must be derived. For point particles that are only translating, these issues do not arise. Here both types of SDE are equivalent and the concentration is equal to the probability density. In this case, the overdamped Langevin equation is equivalent to the Fokker-Planck equation (\cite{gardiner85, lelievre10}) which is the same as the diffusion equation (in this case also called the Smoluchowski equation).

Derivations of the equivalence for some specific situations are known, for example, the diffusion of a dipole in polar coordinates is given in \cite{coffey04}. The proof in an arbitrary manifold is given by \cite{chirikjian09, chirikjian11}. In this paper, the goal is to show the equivalence for two physically important cases: diffusion of a sphere in $S^2$, and diffusion in $SO(3)$. An example of diffusion of a sphere in $S^2$ is the motion of spherical colloids or Janus particles that have a dipole moment. An example of diffusion of a sphere in $SO(3)$ is a spherical virus protein with several bonding sites on the surface. Many coordinates systems can be used to characterize rotations \cite{goldstein62}, therefore all proofs are done in arbitrary coordinate systems. Examples of typical coordinates such as polar angles and quaternions are also given. In the final part of the paper, the rotational diffusion equations are combined with the translational diffusion, and a SDE and heat equation is given for an arbitrary number of 
particles.

\section{Balance of Momentum}
Consider the balance of angular momentum for a spherical particle rotating about its centroid that experiences viscous drag torque, a conservative torque, and a stochastic torque
\begin{align}
 \textbf{I} \dot{\boldsymbol{\omega}}= \boldsymbol{\tau}^{(v)} + \boldsymbol{\tau}^{(c)} + \boldsymbol{\tau}^{(s)}.  \label{eq:angularMomentum}
\end{align}
Because we are considering a sphere, this equation is valid in both a fixed Eulerian frame, which will be denoted by a superscript $L$, and a Lagrangian frame which rotates with the object, denoted by a superscript $R$ (as in \cite{chirikjian01}). The absence of these superscripts indicates the equation works for both reference frames. Here, we are interested only in spheres, so the rotational inertia $\textbf I = a \textbf 1$ where $\textbf{1}$ is the identity matrix. The interested reader could modify this for non-spherical shapes.

Following the notation of \cite{walter10}, the torques acting on the particle are
 \begin{subequations}
 \begin{align}
  & \boldsymbol{\tau}^{(v)} = -8 \pi \mu R^3 \boldsymbol{\omega} \equiv - \gamma \boldsymbol{\omega} \label{eq:Vtau}\\
 &  \dot{U} = \boldsymbol{\tau}^{(c)} \cdot \boldsymbol{\omega} \label{eq:Utau}\\
  &  \boldsymbol{\tau}^{(s)} = \sqrt{2 kT \gamma} \boldsymbol{\xi}(t) \label{eq:Stau}
 \end{align}
 \end{subequations}
where $\mu$ is the viscosity of the surrounding fluid, $R$ is the radius of the particle, $\omega$ is the rotational velocity, $\gamma$ is a friction coefficient for rotational motion defined by eq.~\ref{eq:Vtau}, $\dot{U}$ is the time derivative of a potential field, $k$ is Boltzmann's constant, $T$ is the temperature, and $\boldsymbol{\xi}(t)$ is a three dimensional white noise vector. Equation \ref{eq:Vtau} assumes the particle rotation is viscous dominated (low Reynolds number $Re=\frac{\rho \omega R^2}{\mu} \ll 1$) in a surrounding Newtonian fluid. The definition of the conservative torque in eq.~\ref{eq:Utau} is standard \cite{goldstein62}. The magnitude of the stochastic term in eq.~\ref{eq:Stau} is $<\xi(t) \xi(t')> = \delta(t-t')$ where the ensamble average is defined in ref~\cite{coffey04}. The ensamble averaged angular velocity at large times from eq.~\ref{eq:angularMomentum} (in the absence of a potential term) is then equal to the equilibrium value based on the equipartition of energy $<\omega^2>= \frac{3kT}{a}$.

When $\boldsymbol{\tau}^{(s)} \sim \boldsymbol{\tau}^{(c)} \ll \boldsymbol{\tau}^{(v)}$, the inertial terms in eq.~\ref{eq:angularMomentum} are negligible \cite{lelievre10}. The overdamped limit of the conservation of angular momentum is
\begin{align}
 \gamma \boldsymbol{\omega}= \boldsymbol{\tau}^{(c)} + \sqrt{2 kT \gamma} \boldsymbol{\xi}(t). \label{eq:overdamped}
 \end{align}
This overdamped equation will be used for the remainder of the paper.

\subsection{Scaling} \label{sec:Scaling}
Equations~\ref{eq:overdamped} and \ref{eq:Utau} are scaled by the minimum potential energy $U_0 = |\min(U)|$. Thus define  $\tilde{\boldsymbol{\omega}} = \frac{\gamma \boldsymbol{\omega}}{U_0}$, $\tilde{t} = t U_0/\gamma$, $\tilde{U} = U/U_0$, $\tilde{\boldsymbol{\xi}} = \boldsymbol{\xi} \sqrt{\gamma/U_0}$, and $\tilde{\beta}^{-1} = kT/U_0$. Then eq.~\ref{eq:overdamped} becomes
\begin{align}
 \tilde{\boldsymbol{\omega}}=  \tilde{\boldsymbol{\tau}}^{(c)} + \sqrt{2 \tilde{\beta}^{-1}} \tilde{\boldsymbol{\xi}}(\tilde{t}) \label{eq:overdampedDimensionless}
 \end{align}
and eq.~\ref{eq:Utau} becomes
 \begin{align}
  \dot{\tilde{\textbf U}} = -\tilde{\boldsymbol{\tau}}^{(c)} \cdot \tilde{\boldsymbol{\omega}} \label{eq:Uomegatau}
 \end{align}
For convenience, the tildes will now be dropped even though dimensionless variables will be used for the remainder of the paper. Index notation will be used. Unless specified otherwise, double Latin indices are summed from one to three, and double Greek indices are summed from one to two. The Levi-Civita permutation symbol is denoted as $\epsilon_{ijk}$ and the symbol for the  Kronecker delta is $\delta_{ij}$.

\section{Motion in $S^2$}
Consider the diffusion of a molecule on the surface of a sphere, with a unit vector $r$ pointing to the molecule. Equivalently, consider the rotational diffusion of a molecule which is spherical but whose orientation can be specified by a single vector, such as a spherical particle with an embedded magnetic dipole. In the Eulerian frame, the unit vector $\textbf r^L$ can be parameterized by two variables
\begin{align}
 r^L_i = r^L_i(\eta_1, \eta_2).
\end{align}
The change in the position vector, or the change in the parameterized variables $\eta$, are related to the angular velocity by
\begin{align}
 \dot{r}^L_i = \pd{r^L_i}{\eta_\alpha} \dot{\eta}_\alpha = \epsilon_{ijk} \omega^L_j r^L_k.
\end{align}
This equation is multiplied by $\pd{r^L_i}{\eta_\gamma}$. Defining the (symmetric) metric tensor as $g_{\gamma \alpha} = \pd{r^L_i}{\eta_\gamma} \pd{r^L_i}{\eta_\alpha}$ and its inverse as $g^{-1}_{\alpha \beta}$ such that $g_{\gamma \alpha} g^{-1}_{\alpha \beta} = \delta_{\gamma \beta}$, the metric tensor can be moved to the right side of the equation, then
\begin{align}
 \dot{\eta}_\beta = \epsilon_{ijk} \omega^L_j r^L_k \pd{r_i^L}{\eta_\gamma} g^{-1}_{\beta \gamma}. \label{eq:etadotbeta}
\end{align}
The torque can be found by using the chain rule on $U(\eta_1,\eta_2)$ and substituting eq.~\ref{eq:etadotbeta} for $\dot{\eta}_\beta$
\begin{align}
\dot{U} =  \pd{U}{\eta_\beta} \dot{\eta}_\beta = - \left[-\pd{U}{\eta_\beta} \epsilon_{ijk} r^L_k \pd{r_i^L}{\eta_\gamma} g^{-1}_{\beta \gamma}\right] \omega^L_j
\end{align}
the Eulerian torque is the quantity in brackets (by eq.~\ref{eq:Uomegatau}).

Substituting in the expression for the angular velocity and multiplying by $dt$ and renaming indices in the torque yields
\begin{align}
 d \eta_\beta = \epsilon_{ijk} r^L_k \pd{r^L_i}{\eta_\gamma} g^{-1}_{\beta \gamma} \left(-\pd{U}{\eta_\alpha} \epsilon_{mjl} r^L_l \pd{r_m^L}{\eta_\kappa} g^{-1}_{\alpha \kappa} dt + \sqrt{2\beta^{-1}} \circ dW_j \right), \label{SDEetabetaLong}
\end{align}
where $\circ dW$ denotes a Stratonovich stochastic differential equation (see \cite{evans01}). The Stratonovich interpretation is used when the underlying noise is smooth \cite{evans01}, as it is in this case due to the overdamped approximation. The SDE~\ref{SDEetabetaLong} is re-labeled as
\begin{align}
 d \eta_\beta = a_\beta dt + b_{\beta j} \circ dW_j \label{SDEetabetaShort}
\end{align}
for simplicity. The expression for $a_\beta$ can be reduced using $r^L_\alpha \pd{r^L_i}{\eta_\alpha} =0$ (since $r^L_i r^L_i=1$) and identities of the Levi-Civita symbol
\begin{align}
 &a_\beta = -\epsilon_{ijk} \epsilon_{mjl} r^L_k r^L_l \pd{r^L_i}{\eta_\gamma} \pd{r^L_m}{\eta_\kappa} \pd{U}{\eta_\alpha} g^{-1}_{\kappa \alpha} g^{-1}_{\kappa \alpha} g^{-1}_{\beta \gamma}\\
& a_\beta = -\pd{U}{\eta_\gamma} g^{-1}_{\beta \gamma}.
 \end{align}
The expression for $b_{\beta j}$ is
 \begin{align}
  b_{\beta j} = \epsilon_{ijk} r^L_k \pd{r^L_i}{\eta_\gamma} g^{-1}_{\beta \gamma} \sqrt{2\beta^{-1}}.
 \end{align}
The Stratonovich interpretation can be verified by comparing the Ito and Stratonovich calculus on the function $f=r^L_i r^L_i$ and checking that the change in $f$ is zero. For Ito calculus, the change in $f$ is $df= \pd{f}{\eta_\alpha} d\eta_\alpha + \frac{1}{2} \frac{\partial^2 f}{\partial \eta_\alpha \partial_\beta} b_{\alpha i} b_{\beta i} dt$. The first term is zero, but the second term is not. For Stratonovich calculus, the second term is not present, and the vector $\textbf r^L$ stays on the sphere.

The Fokker-Planck equation  for the Stratonovich SDE \cite{chirikjian09, gardiner85} is
\begin{align}
 \dot{p} = -\pd{}{\eta_\beta}\left(a_\beta p - \frac{1}{2} b_{\beta j} \pd{}{\eta_\alpha}(b_{\alpha j} p ) \right).
\end{align}
Here $p$ represents the probability of being in a given area $d\eta_1 d\eta_2$. In physics, the convention is to specify a concentration ($c$), which is the probability of being in a certain area $\sqrt{g} d\eta_1 d\eta_2$. A constant concentration implies a uniform distribution about the sphere. The conversion between these two is given by
\begin{align}
 p = \sqrt{g} c
\end{align}
where $g$ is the determinant of the metric tensor. The Fokker-Planck equation then becomes
\begin{align}
 \dot{c} = -\frac{1}{\sqrt{g}} \pd{}{\eta_\beta}\left(a_\beta \sqrt{g} c - \frac{1}{2} b_{\beta j} \pd{}{\eta_\alpha}(b_{\alpha j} \sqrt{g} c ) \right).
\end{align}
as derived in \cite{chirikjian09}. This is related to the heat equation using a Piola-type identity $\pd{}{\eta_\alpha} \left(\sqrt{g} b_{\alpha j} \right)=0$, proven in appendix \ref{sec:PiolaS2}. Algebraic manipulation shows $b_{\beta j} b_{\alpha j} = 2 \beta g^{-1}_{\alpha \beta}$ and thus the heat equation
\begin{align}
 \dot{c} = \frac{1}{\sqrt{g}} \pd{}{\eta_\beta} \left( g^{-1}_{\beta \gamma} \sqrt{g} c \pd{U}{\eta_\gamma} + \beta^{-1} g^{-1}_{\beta \gamma} \sqrt{g} \pd{c}{\eta_\gamma} \right)
\end{align}
is obtained. It is readily verified that the equilibrium solution to this is $c_{eq} = \exp(-\beta U)$. This equation is the diffusion equation on a sphere obtained by Perrin \cite{perrin34} in the absence of the potential term. This can also be written in terms of the divergence and gradient operators \cite{kreyszig91} on a surface
\begin{align}
 \dot{c} = \nabla_s \cdot (c \nabla_s U) + \beta^{-1} \triangle_s c.
\end{align}

\subsection{Polar Coordinates}
This section is for the reader interested in applying the results in $S^2$ to the case of polar coordinates. The convention for the position vector $r^L = (\sin \theta \cos \phi, \sin \theta \sin \phi, \cos \theta)$ is used. The SDE in these coordinates is
\begin{subequations} 
\begin{align}
& d\theta = \pd{U}{\theta} dt + \sqrt{2\beta^{-1}} \circ dV_1 \\
& d\phi = \frac{1}{\sin^2(\theta)} \pd{U}{\phi} dt + \frac{\sqrt{2\beta^{-1}}}{\sin(\theta)} \circ  dV_2
\end{align} \label{eq:sphericalcoordinatesSDE}
\end{subequations}
where
\begin{subequations}
 \begin{align}
& dV_1 = -\sin(\phi)dW_1+\cos(\phi)dW_2\\
& dV_2 = -\cos(\phi) \cos(\theta) dW_1 - \sin(\phi) \cos(\theta) dW_2+\sin(\theta) dW_3.
 \end{align}\label{eq:dWdV}
\end{subequations} 
The Ito form of this equation is
\begin{subequations} 
\begin{align}
& d\theta = \left(\pd{U}{\theta}+ \beta^{-1} \cot(\theta) \right) dt + \sqrt{2\beta^{-1}} dV_1 \label{eq:dtheta}\\
& d\phi = \frac{1}{\sin^2(\theta)} \pd{U}{\phi} dt + \frac{\sqrt{2\beta^{-1}}}{\sin(\theta)}  dV_2.
\end{align} \label{eq:sphericalcoordinatesSDEito}
\end{subequations}
$V_1$ and $V_2$ can be computed by eq~\ref{eq:dWdV} or can be considered to be independent Brownian motions (\cite{brillinger12, chirikjian09}). For both sets of equations~\ref{eq:sphericalcoordinatesSDE} and \ref{eq:sphericalcoordinatesSDEito}, the probability density equation for the concentration is the heat equation on the sphere
\begin{align}
 \dot{c} = & \frac{1}{\sin(\theta)} \pd{}{\theta} \left( \sin(\theta) c \pd{U}{\theta} \right) + \frac{1}{\sin{\theta}} \pd{}{\phi} \left( \frac{1}{\sin(\theta)}c\pd{U}{\phi}\right) + \nonumber \\
 & \beta^{-1} \left( \frac{1}{\sin(\theta)} \pd{}{\theta} \left(\sin(\theta) \pd{c}{\theta}\right)+\frac{1}{\sin^2(\theta)}\frac{\partial^2 c}{\partial \phi^2}\right). \label{eq:HeatPolar}
\end{align}

\begin{figure}[!ht]
\begin{tabular}{cc}
\includegraphics [width=5 cm, clip]{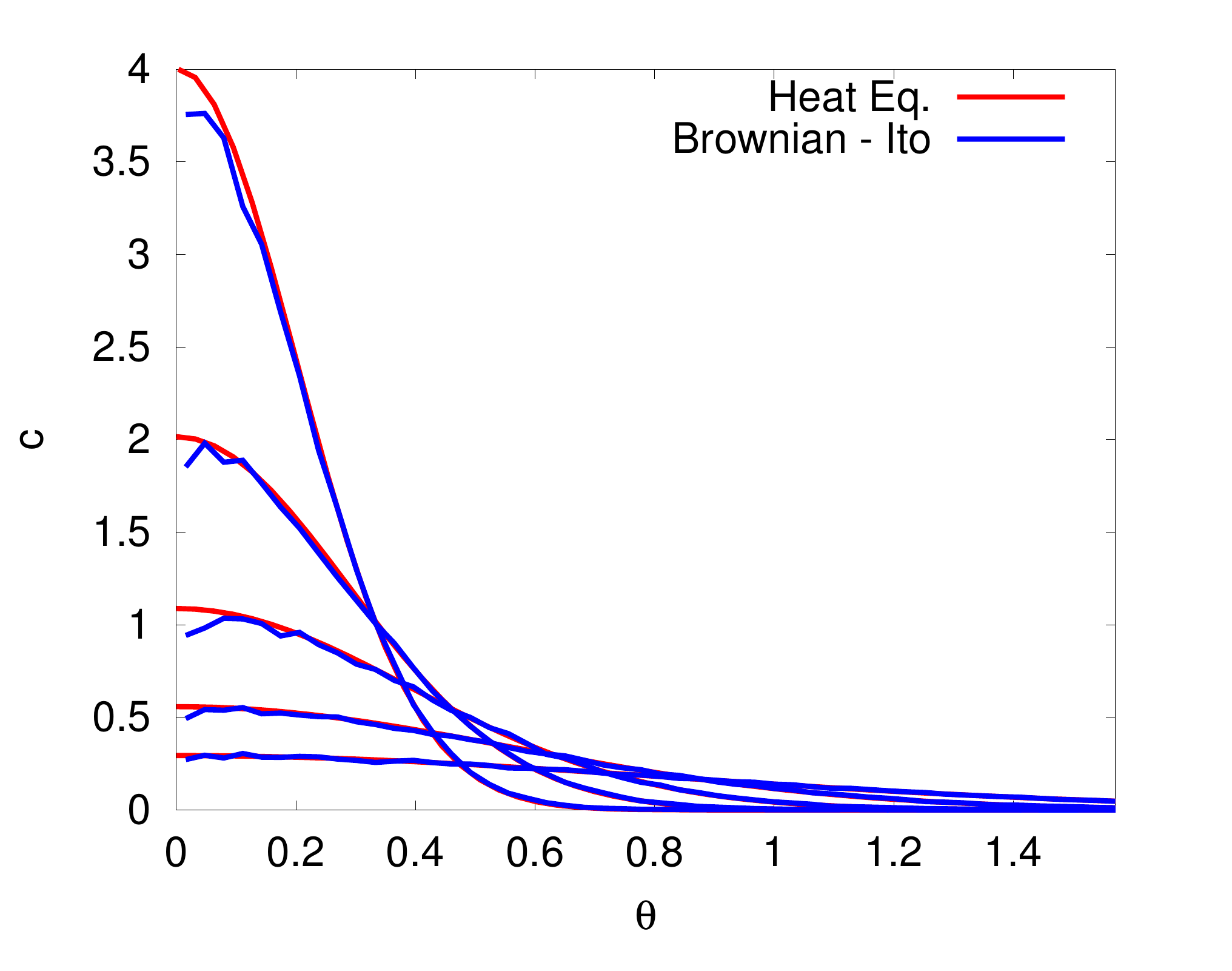}&
  \includegraphics [width=5 cm, clip]{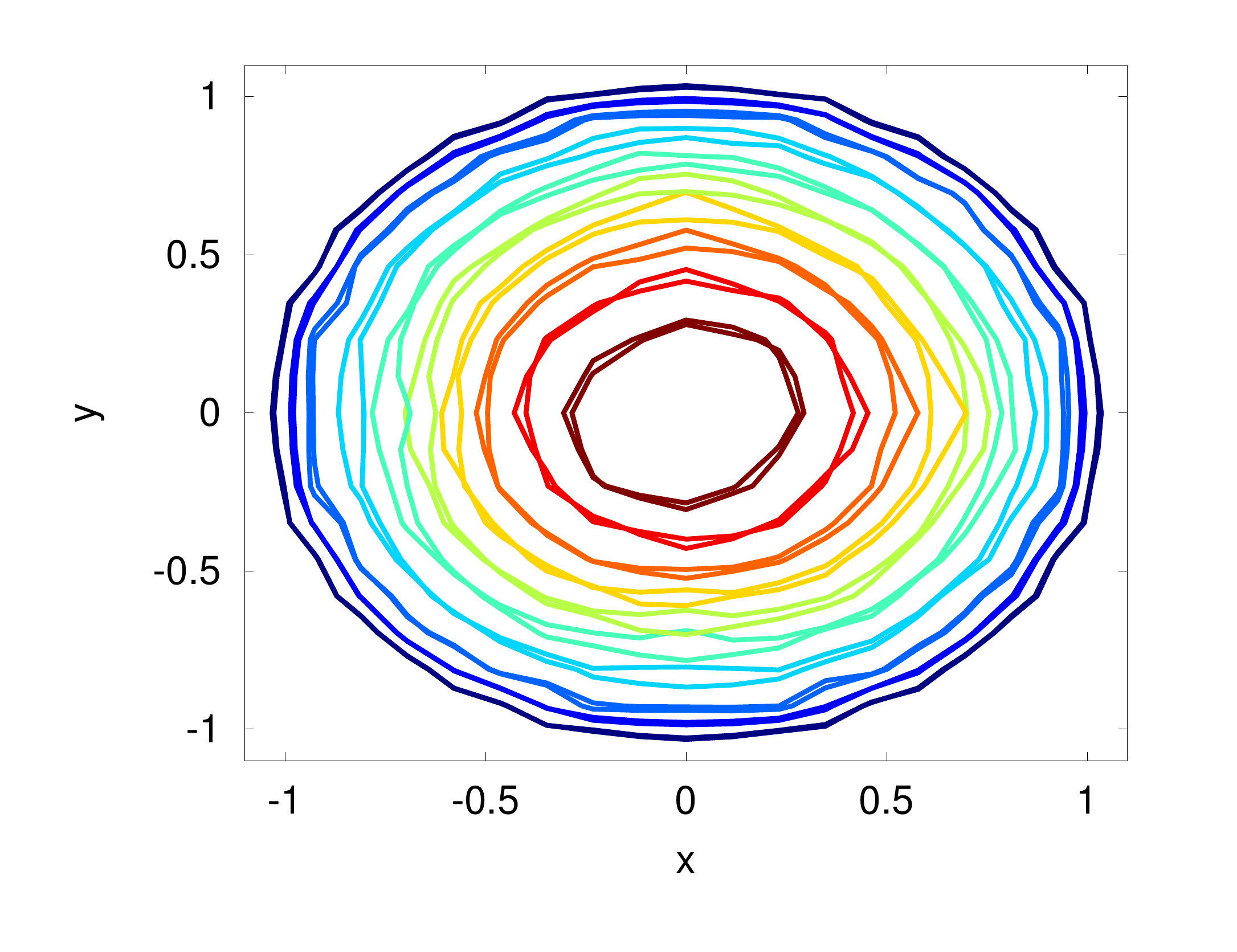}
  \end{tabular}
\caption{Left: Concentration as a function of polar angle at times $0.04$, $0.08$, $0.15$, $0.30$, and $0.60$ with $\beta =1$ for the solution of the heat equation compared to an average of $100,000$ trials of eq.~\ref{eq:dtheta} with time step $dt=10^{-5}$. Right: Contour plots of Brownian diffusion on a sphere. The starting point is $\theta=\pi/4$. The contours show the concentration of $1,000,000$ particle positions at the end of $t=0.30$ ($\beta=1$) when viewing the sphere with the initial condition at the center. The contour plot reflected about the diagonal is also plotted, to verify the symmetry.}\label{fig:Discrete}
\end{figure}

The SDE~\ref{eq:sphericalcoordinatesSDEito} and the heat equation (eq.~\ref{eq:HeatPolar}) were compared against one another in the absence of a potential term using numerical simulation in figure~\ref{fig:Discrete}. The numerical integration of eqs.~\ref{eq:sphericalcoordinatesSDEito} was performed using the Euler-Maruyama method. The parameterization of the unit vector in terms of $\theta$ and $\phi$ has the advantage over an SDE for $d\textbf{r}$ because there is no need to project back to the surface of the sphere. The drawback of the SDE parameterization is that if $\theta$ randomly jumps too close to $0$ or $\pi$, the cotangent term diverges; therefore these points were moved away from the poles a small set distance. If a point happens to go beyond $0$ or $\pi$ it is reflected back into the domain. The heat equation (eq.~\ref{eq:HeatPolar}) was solved by method of Legendre polynomials. Figure~\ref{fig:Discrete} shows good agreement for the time evolution for a particle starting at the pole $\theta=0$. This requires only equation~\ref{eq:dtheta}. To compare both the equations of the SDE system~\ref{eq:sphericalcoordinatesSDEito}, the right diagram in figure~\ref{fig:Discrete} shows the average distribution viewed from above the point $\theta=\pi/4$. The spherical symmetry demonstrates that there is no prefered direction in the SDE. 

When the Euler-Maruyama method was used to integrate the Strantonovich SDE~\ref{eq:sphericalcoordinatesSDE} then artificially high concentrations occurred around the poles. This could be seen in disagreements between the heat equation and the SDE simulations in a figure like the one in fig~\ref{fig:Discrete} left, as well as a lack of symmetry in a figure of the type in fig~\ref{fig:Discrete}. One way to interpret this disagreement is that Strantonovich noise is smooth and noise in numerical simulations are discontinuous.

\section{Motion in $SO(3)$}
Consider a spherical particle which has several bonding sites which do not affect the drag. One way to describe the orientation of this particle is an orthogonal rotation matrix $\textbf R$. The change in the rotation matrix in the Lagrangian coordinate system $\textbf R ^R$ is related to the angular velocity by
\begin{align}
 \dot{R}^R_{lk} = R^R_{lj} \epsilon_{ijk} \omega^R_i \label{eq:Romega}\\
 \omega^R_i = \frac{1}{2} \epsilon_{ijk} R^R_{lj} \dot{R}^R_{lk}.
\end{align}
Due to the orthogonality condition, any rotation can be parameterized by three variables $ R_{ij}= R_{ij}(\eta_1,\eta_2,\eta_3)$; by the chain rule
\begin{align}
\omega^R_i = S_{in} \dot{\eta}_n \label{eq:omegaSetadot}
\end{align}
where
\begin{align}
S_{in} = \frac{1}{2} \epsilon_{ijk} R^R_{lj} \pd{R^R_{lk}}{\eta_n}. \label{eq:defineS}
\end{align}
The matrix $\textbf S$ is called either the angular velocity structure matrix \cite{walter10} or the right Jacobian \cite{chirikjian01}. Chirikjian \cite{chirikjian01} also defines a left Jacobian, which appears in the Eulerian frame.

The expression for the conservative torque is obtained by plugging eq. \ref{eq:omegaSetadot} into eq. \ref{eq:Uomegatau}. Multiplying by the inverse of $\textbf S$,
\begin{align}
\tau_i^{(c)^R} = -S_{si}^{-1} \pd{U}{\eta_s} \label{eq:tauSU}
\end{align}
Combining eqs.~\ref{eq:omegaSetadot}, \ref{eq:tauSU} and \ref{eq:overdampedDimensionless} and multiplying by $\textbf S^{-1} dt$ gives the SDE for the rotation coordinates
\begin{align}
 d \eta_n = -S^{-1}_{ni} S^{-1}_{si} \pd{U}{\eta_s} dt + \sqrt{2\beta^{-1}} S^{-1}_{ni} \circ dW_i \label{eq:SDEetaR}
\end{align}
where the $\circ dW_i$ denotes a Stratonovich noise term. The underlying noise is smooth and it is verified that $df=0$ where $f=R_{ij} R_{ij}$ only for the Stratonovich calculus. 

As in the previous section, a Fokker-Planck equation for the evolution of the concentration can be obtained from SDE~\ref{eq:SDEetaR}. On the $SO(3)$ manifold, the metric tensor is $\textbf{G} = \textbf{S}^T \textbf{S}$, thus the determinant $\sqrt{G} = \det(\textbf S)$ plays the same roll as $\sqrt{g}$ (see Chirikjian \cite{chirikjian09}). The concentration in $SO(3)$ is defined as $p = \sqrt{G} c$. The evolution of the concentration is
\begin{align}
 \dot{c} = \frac{1}{\sqrt{G}} \pd{}{\eta_n}\left(S^{-1}_{ni} S^{-1}_{si} \pd{U}{\eta_s} \sqrt{G} c + \beta^{-1} S^{-1}_{ni}  \pd{}{\eta_s}\left(S^{-1}_{si} \sqrt{G} c \right)\right).
\end{align}
From the Piola-type identity (proven in Appendix \ref{sec:PiolaSO3}) $\pd{}{\eta_l}\left(S^{-1}_{li} \sqrt{G} \right) =0$
\begin{align}
 \dot{c} = \frac{1}{\sqrt{G}} \pd{}{\eta_n}\left(S^{-1}_{ni} S^{-1}_{si} \pd{U}{\eta_s} \sqrt{G} c + \beta^{-1} S^{-1}_{ni} S^{-1}_{si} \sqrt{G} \pd{c}{\eta_s} \right).
\end{align}
By definition of $G$, this can be expressed as
\begin{align}
 \dot{c} = \frac{1}{\sqrt{G}} \pd{}{\eta_n} \left( G^{-1}_{ns} \sqrt{G} c \pd{U}{\eta_s} + \beta^{-1} G^{-1}_{ns} \sqrt{G} \pd{c}{\eta_s} \right). \label{eq:HeatSO3}
\end{align}
It is readily verified that the equilibrium solution to this is $c_{eq} = \exp(-\beta U)$. Equation~\ref{eq:HeatSO3} is the diffusion equation \cite{perrin34} of a rigid body with a potential term. This can also be written in terms of the divergence and gradient operators on $SO(3)$ (see \cite{chirikjian09}) as
\begin{align}
 \dot{c} = \nabla_S \cdot (c \nabla_S U) + \beta^{-1} \triangle_S c
\end{align}

\subsection{Quaternions}
Numerical simulation of diffusion in $SO(3)$ can be tracked with quaternions \cite{allen89}. The quaternion is of unit length ($q_i q_i = 1$ where $i=1,2,3,4$). The rotation matrix is related to the quaternion by
\begin{align}
 \textbf{R}=\left[ \begin{array}{rrr}
q_1^2+q_2^2-q_3^2-q_4^2 & 2 (q_2 q_3 +q_1 q_4) &  2 (q_2 q_4 - q_1  q_3 )\\
    2 ( q_2  q_3 - q_1  q_4 )&  q_1^2- q_2^2+ q_3^2- q_4^2 & 2 ( q_3  q_4 + q_1  q_2 )\\
    2 ( q_2  q_4 + q_1  q_3 )& 2(q_3  q_4 - q_1  q_2 ) &  q_1^2- q_2^2- q_3^2+ q_4^2\\                    
                   \end{array} \right]
\end{align}
The Stratonovich SDE governing the quaternion $q$ is
\begin{align}
dq_i = Q_{ij}\left(\tau^{(c)}_j dt + \sqrt{2 \beta^{-1}} \circ dW_j \right)
\end{align}
where
\begin{align}
 \textbf{Q} = \left[ \begin{array}{rrr}
               -q_2 & -q_3 & -q_4\\
               q_1 & -q_4 & q_3\\
               q_4 & q_1 & -q_2\\
               -q_3 & q_2 & q_1\\
              \end{array} \right]      
\end{align}
The Ito SDE is
\begin{align}
dq_i = Q_{ij}\left(\tau^{(c)}_j dt + \sqrt{2 \beta^{-1}} dW_j \right) -3 q_i dt 
\end{align}
Unlike the simulations in polar coordinates, the simulations of quaternions with Ito and Stratonovich interpretations are the same in the limit of small $dt$. After each time iteration $q$ must be normalized, therefore any addition of a term of the form $q_i$ causes only an insignificant difference.

\section{Multiple particles with translation and rotation}
The equations in the previous sections can be generalized to model the behavior of several particles which both rotate and diffuse. The translational friction coefficient for a sphere is $\gamma_T = 6 \pi \mu R$, and this enters the drag and stochastic terms in an analagous way as defined in equation~\ref{eq:angularMomentum} so that the varience of the velocity distribution matches that from the equipartition of energy. The distance $x$ has been scaled by the radius, and the scaling of time and noise terms are the same as in section~\ref{sec:Scaling}. For particle $n$ where $n=1,2,...,N$, the SDE for both translation and rotation in $SO(3)$  is
\begin{align}
 & dx^n_i = -D_R \pd{U}{x_i^n} dt + \sqrt{2 D_R \beta^{-1}} \circ dW'^n_i\\
 & d \eta^n_j = -S^{-1}_{ji} S^{-1}_{si} \pd{U}{\eta^n_s} dt + \sqrt{2 \beta^{-1}} S^{-1}_{ji} \circ dW^n_i \label{eq:TRrotational}
\end{align}
where $D_R = \frac{\gamma}{\gamma_T R^2}$ is a ratio rotational to translational drag. For spheres $D_R = 4/3$. All $dW$ and $dW'$ terms are independent, ie. $dW_i dW_j = \delta{ij} dt$ and $dW'_i dW_j = 0$. The translational diffusion is the same for both Ito and Stratonovich interpretations. Generalization to $S^2$ would simply replace eq.~\ref{eq:TRrotational} with eq.~\ref{SDEetabetaShort}.

The resulting Fokker-Planck equation for the evolution of the concentration of the $N$ particles ($c=c(\textbf{x}^1,\boldsymbol{\eta}^1,...,\textbf{x}^N,\boldsymbol{\eta}^N$) is
\begin{align}
 \dot{c} =& \sum_{n=1}^N D_R\left( \pd{}{x^n_i} \left(c \pd{U}{x^n_i}\right) + \beta^{-1} \frac{\partial^2 c}{\partial x^n_i \partial x^n_i} \right) + \nonumber\\
 &\frac{1}{\sqrt{G^n}} \pd{}{\eta_i^n} \left((G_{is}^n)^{-1} \sqrt{G^n} c \pd{U}{\eta_s^n}\right) + \beta^{-1} \frac{1}{\sqrt{G^n}} \pd{}{\eta_i^n} \left((G_{is}^n)^{-1} \sqrt{G^n} \pd{c}{\eta_s^n}\right)
\end{align}
or, using the divergence, gradient, and Laplace-Beltrami operators defined above
\begin{align}
 \dot{c} = \sum_{n=1}^N  D_R \left(\nabla^n \cdot (c \nabla^n U) + \beta^{-1} \triangle^n c \right) + \nabla_S^n \cdot (c \nabla_S^n U) + \beta^{-1} \triangle_S^n c.
\end{align}
The equilibrium concentration is $c=\exp(-\beta U)$

\section{Conclusions}
The stochastic differential equation that governs the overdamped rotation of spherical particles was shown to ensemble average to a concentration distribution that evolves by the heat equation. Although one might expect this should be true, it involved three subtle points.

First, the overdamped equation is an SDE of Stratonovich, and not Ito, form. The physical interpretation here is that the overdamped approximation smooths out the the white noise, and the angular velocity must be a continuous function. For stochastic rotational motion that includes inertia or for stochastic translational motion, this this difference is inconsequential \cite{walter10} - the physical interpretation being that the Fokker-Planck equation does not depend on the exact details of the collisions that cause the stochastic forcing. The distinction is only important for some coordinate systems. For rotations modeled using a unit vector in $S^2$ or quaternions in $SO(3)$ this distinction is irrelevant. For polar coordinates, the Strantonovich interpretation governs the behavior, and this must be converted to an Ito equation for numerical simulations.

Second, there is a difference between the probability (obtained by the Fokker-Planck equation) and the concentration (typically used by physicists and engineers). In $S^2$ concentration has a very natural interpretation as a probability per infinitesimal area - and this area can change in size based on the location in space. For $SO(3)$ the conversion between probability and concentration is given by the determinant of the structure factor matrix $S$ (also pointed by \cite{walter10, chirikjian11}). This difference is also reflected in the equilibrium distribution, which for probability distributions is $p=\exp(-\beta U) \det(S)$, whereas for concentration, the more conventional $c=\exp(-\beta U)$ is obtained. 

Third, a Piola-type identity was necessary to simplify the concentration distribution function to obtain the Laplace operator. In $S^2$ identities from differential geometry were used to prove the identity. In $SO(3)$ the identification of skew-symmetric tensors was crucial in proving this identity. 

Many practitioners may be interested in finding trends of systems that contain many particles that both rotate and translate. This paper showed how to generalize the SDEs and obtain the heat equation for translation and rotation. The result is quite a large system of equations, which are efficiently evaluated numerically using GPUs. To gain analytical results, it is desirable to reduce the dimensionality of the heat equation for multiple identical particles. This is an active area of research, where it would be beneficial to improve or replace existing techniques such as BBGKY (see Condiff \cite{Condiff66}).

\section*{Acknowledgment}
This work was supported by an NSF PIRE Award number 0967140. Richard James provided helpful discussions to prove the Piola-type identities.

\appendix
\section{Piola-type identity}
The identity
\begin{align}
\pd{}{\eta_l}\left(M^{-1}_{li} \sqrt{J} \right) =0
\end{align}
for a matrix $M$, which is the gradient of a potential (ie. $\pd{M_{ij}}{\eta_k} = \pd{M_{ik}}{\eta_j}$), and $J=\det(M)$ is called the Piola identity \cite{hughes94}. However, the matrix $S$ does not satisfy $\pd{S_{ij}}{\eta_k} = \pd{S_{ik}}{\eta_j}$, therefore it cannot be a potential, and the Piola identity cannot be applied. The derivation of $\pd{}{\eta_l}\left(S^{-1}_{li} \sqrt{G} \right) =0$ follows. For the non-square matrix $b_{\alpha j}$ the identity is also generalized below. This identity is true in general for the Jacobian of a unimodular Lie group (see \cite{chirikjian11}), however, the proof below does not require any properties of the group.

\subsection{Piola in $S^2$} \label{sec:PiolaS2}
The Piola identity on the surface of a sphere
\begin{align}
 \pd{}{\eta_\alpha} \left(\sqrt{g} b_{\alpha j} \right)=0
\end{align}
is proven by first applying the derivative to each of the components, using typical simplifications \cite{kreyszig91}
\begin{align}
 \pd{}{\eta_\alpha} \left(\sqrt{g} b_{\alpha j} \right)=& \sqrt{g} \Gamma_{\kappa \alpha \gamma} g^{-1}_{\kappa \gamma} \epsilon_{ijk} r^L_k \pd{r^L_k}{\eta_\gamma} g^{-1}_{\alpha \gamma} +\sqrt{g} \epsilon_{ijk} \pd{r^L_k}{\eta_\alpha} \pd{r^L_i}{\eta_\gamma} g^{-1}_{\alpha \gamma} \nonumber \\
 & + \sqrt{g} \epsilon_{ijk} r^L_k \frac{\partial^2 r^L_i}{\partial \eta_\gamma \partial \eta_\alpha} g^{-1}_{\alpha \gamma} +\sqrt{g} \epsilon_{ijk} r^L_k \pd{r^L_i}{\eta_\gamma} \pd{g^{-1}_{\alpha \gamma}}{\eta_\alpha}
\end{align}
The first term expresses the derivative of the square root of the metric tensor in terms of the Christoffel symbol $\Gamma$. The second term is zero by permutation of indices. The third and fourth terms can be written in terms of Christoffel symbols as well, to obtain
\begin{align}
 \pd{}{\eta_\alpha} \left(\sqrt{g} b_{\alpha j} \right)=
 & \sqrt{g} \Gamma_{\kappa \alpha \gamma} g^{-1}_{\kappa \gamma} \epsilon_{ijk} r^L_k \pd{r^L_i}{\eta_\gamma} g^{-1}_{\alpha \gamma} \nonumber \\
 & + \sqrt{g} \epsilon_{ijk} r^L_k \Gamma_{\gamma \alpha \kappa} g^{-1}_{\kappa \beta} \pd{r^L_i}{\eta_\beta} g^{-1}_{\alpha \gamma} 
 - \sqrt{g} \epsilon_{ijk} r^L_k \pd{r^L_i}{\eta_\gamma} g^{-1}_{\alpha \gamma} g^{-1}_{\kappa \beta} \left(\Gamma_{\beta \alpha \gamma} + \Gamma_{\gamma \alpha \beta} \right)
\end{align}
The Christoffel symbol is symmetric in the first two indices, and the metric tensor is also symmetric, therefore this expression is zero.

\subsection{Piola in $SO(3)$} \label{sec:PiolaSO3}
 By definition of the cofactor matrix $\textbf C$ of $\textbf S$,
\begin{align}
 \pd{}{\eta_l}\left(S^{-1}_{li} \sqrt{G} \right) = \pd{C_{il}}{\eta_l}
\end{align}
The cofactor matrix can be expressed in terms of permutation symbols (see \cite{bhattacharya03})
\begin{align}
 \pd{}{\eta_l}\left(S^{-1}_{li} \sqrt{G} \right) = \frac{1}{2} \pd{}{\eta_l} \left( \epsilon_{ikj} \epsilon_{lpr} S_{kp} S_{jr} \right)
\end{align}
Inserting the definition of $\textbf S$ from eq.~\ref{eq:defineS}, and expanding permutation symbols
\begin{align}
 \pd{}{\eta_l}\left(S^{-1}_{li} \sqrt{G} \right) = \frac{1}{4} \pd{}{\eta_l} \left( \epsilon_{lpr} \epsilon_{jdf} R_{cj} \pd{R_{ci}}{\eta_p} R_{gd} \pd{R_{gf}}{\eta_r} \right)
\end{align}
The derivative with respect to $\eta_l$ is taken. Simplifications are made using the identity $R_{nf} \epsilon_{ncg} = \epsilon_{jdf} R_{cj} R_{gd}$ (obtained by writing the inverse of $\textbf{R}$ as its transpose and also using the cofactor expansion from \cite{bhattacharya03} where the determinant is one). The expression
\begin{align}
 \pd{}{\eta_l}\left(S^{-1}_{li} \sqrt{G} \right) = \frac{1}{4} \epsilon_{lpr} \epsilon_{jdk} R_{nk} \pd{R_{nf}}{\eta_l} R_{cj} \pd{R_{ci}}{\eta_p} R_{gd} \pd{R_{gf}}{\eta_r}
\end{align}
is obtained. From eq.~\ref{eq:Romega}, $\textbf{R}^T \dot{\textbf{R}}$ is skew-symmetric. Because the functional form of the rotation matrix is not defined, there are no restrictions on the relation between $\eta$ and time. Therefore, for any $\eta_l$, the matrix $R_{nk} \pd{R_{nf}}{\eta_l}$ is skew-symmetric, and can be expressed as 
\begin{align}
K^l_a \epsilon_{akf} = R_{nk} \pd{R_{nf}}{\eta_l}
\end{align}
where $K^l_a$ is an unknown matrix (superscripts are used only to clarify there are $l$ skew-symmetric matrices). Therefore
\begin{align}
 \pd{}{\eta_l}\left(S^{-1}_{li} \sqrt{G} \right) = \frac{1}{4} \pd{}{\eta_l} \left( \epsilon_{lpr} \epsilon_{jdk} K^l_a \epsilon_{akf} K^p_q \epsilon_{qji} K^r_b \epsilon_{bdf}\right).
\end{align}
By applying simplifications due to sums of permutation symbols, the right hand side becomes
\begin{align}
 \pd{}{\eta_l}\left(S^{-1}_{li} \sqrt{G} \right) = \frac{1}{2} \epsilon_{lpr} K^l_q K^p_q K^r_i 
\end{align}
which is zero, by exchange of indices $l$ and $p$.

\bibliographystyle{plain}

\end{document}